\documentclass[journal]{IEEEtran}
\usepackage[utf8x]{inputenc} 
\usepackage{textcomp}
\usepackage{cite}
\usepackage{amsmath,amssymb,amsfonts}
\usepackage{algorithmic}
\usepackage{graphicx}
\usepackage{textcomp}
\usepackage{xcolor}
\usepackage{graphicx,wrapfig,lipsum}
\usepackage{color}
\usepackage{comment}

\newcommand{\blue}{\color{black}}
\newcommand{\green}{\color{black}}
\usepackage[ruled,vlined]{algorithm2e}
\def\BibTeX{{\rm B\kern-.05em{\sc i\kern-.025em b}\kern-.08em
    T\kern-.1667em\lower.7ex\hbox{E}\kern-.125emX}}
\begin{document}

\title{ Physical \textit{Time-Varying Transfer Functions} as Generic Low-Overhead Power-SCA Countermeasure 
}

\author{\IEEEauthorblockN{Archisman Ghosh, Debayan Das and Shreyas Sen}\\

\IEEEauthorblockA{\textit{School of Electrical and Computer Engineering, Purdue University, IN, USA}}

\thanks{A. Ghosh, D. Das, and S. Sen are with the School of Electrical and Computer Engineering, Purdue University, West Lafayette, IN, 47906 USA email: (ghosh69@purdue.edu, das60@purdue.edu, shreyas@purdue.edu).}
}

\maketitle

\begin{abstract}
Mathematically-secure cryptographic algorithms leak significant side-channel information through their power supplies when implemented on a physical platform. These side-channel leakages can be exploited by an attacker to extract the secret key of an embedded device. The existing state-of-the-art countermeasures mainly focus on the power balancing, gate-level masking, or signal-to-noise (SNR) reduction using noise injection and signature attenuation, all of which suffer either from the limitations of high power/area overheads, performance degradation or are not synthesizable. In this article, we propose a generic low-overhead digital-friendly power SCA countermeasure utilizing physical Time-Varying Transfer Functions (TVTF) by randomly shuffling distributed switched capacitors to significantly obfuscate the traces in the time domain. System-level simulation results of the TVTF-AES implemented in TSMC 65nm CMOS technology show $>4000\times$ minimum traces to disclosure (MTD) improvement over the unprotected implementation with $\sim1.25\times$ power and $\sim 1.2\times$ area overheads, and without any performance degradation.
\end{abstract}

\begin{IEEEkeywords}
Power Side-Channel Attack, Low-overhead Countermeasure, Physical Obfuscation, Time-varying transfer function, Synthesizable, Generic.
\end{IEEEkeywords}

\section{Introduction}
In today's data-driven internet-connected (IoT) world, security and confidentiality of communication and computing are of utmost importance. To address these needs, various cryptographic algorithms have been proposed till date, which are computationally secure. However, as these algorithms are implemented on a physical substrate, it leaks critical 'side-channel' information in the form of power consumption \cite{kocher_differential_1999}, \cite{brier_optimal_2003}, electromagnetic (EM) emanation \cite{quisquater_electromagnetic_2001}, \cite{gandolfi_electromagnetic_2001}, cache hits/misses \cite{kocher_timing_1996}, \cite{brumley_remote_2003}, and so on. These side-channel leakages can be exploited by attackers to extract the secret key from a cryptographic device. In this article, we focus on the power SCA attack on an AES-128 engine.

\begin{figure}[!t]
\includegraphics[width=\columnwidth]{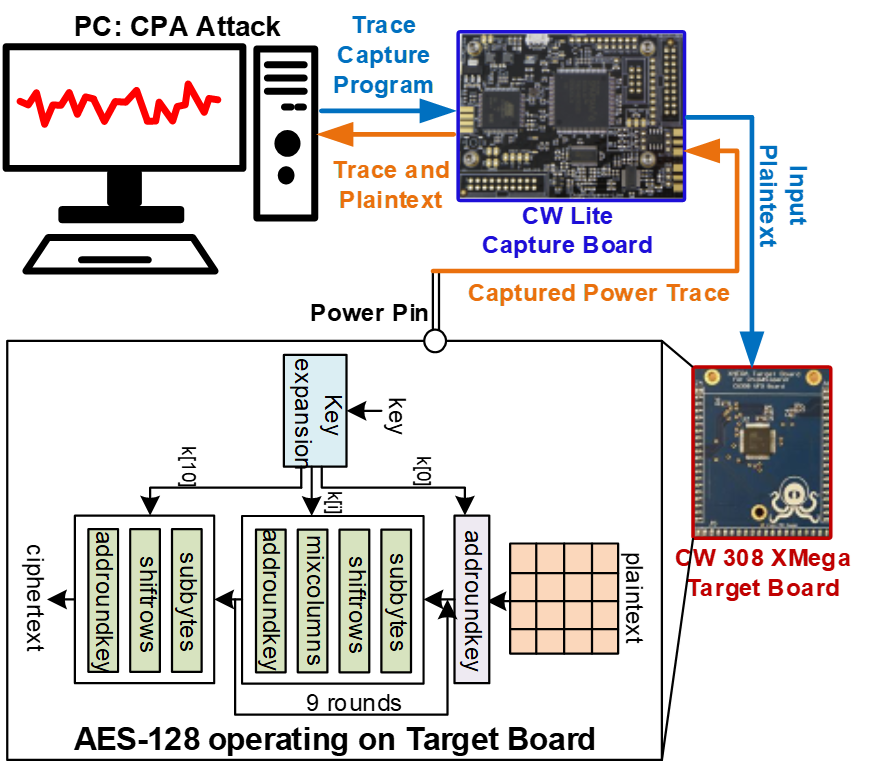}
\caption{\label{fig:setup} Power SCA attack setup using Chipwhisperer on an 8-bit Atmega microcontroller running AES-128 encryption.}
\label{intro}
\end{figure}

Power analysis attack is one of the most common side-channel attacks on embedded systems. The time-complexity of breaking an AES-128 engine is reduced from $2^{128}$ for a brute-force attack to $2^{12}$ for a power SCA attack, as the key search space reduces to $2^8 = 256$ possibilities for each of the 16 key bytes. 

Power SCA is performed by measuring the power consumption of a target device during the encryption phase. Every captured trace is synchronized with a chosen plaintext (PT) or a known ciphertext (CT). The attacker can either feed chosen PTs to the target device or record the output CT, while capturing the power traces. Once the traces are collected, a differential/correlational power analysis (DPA/CPA) attack \cite{kocher_differential_1999}, \cite{brier_optimal_2003} is performed using a hamming weight (HW) or a hamming distance (HD) model. The HW leakage model considers the number of ones on the data bus during a switching activity, while HD model takes into account the number of bits switching from one state to the next. HW models are useful for software crypto implementations on various microcontrollers (as in this work), while HD models are typically used for attacks on hardware crypto implementations where the operations are highly parallelized and the same register is updated each clock cycle to store the updated state. 

Fig.~\ref{intro} shows an overview of the power SCA attack set-up using the Chipwhisperer platform \cite{oflynn_chipwhisperer:_2014}. Traces are collected from an unprotected software AES-128 engine running on an 8-bit Atmega microcontroller for varying chosen input plaintexts {\blue which has been provided by Chipwhisperer board to the micro-controller}. The traces from the Xmega target board is transferred to the PC, where a CPA attack is performed to extract the secret 128-bit key. In this work, we use the HW model for the leakage from the microcontroller device and target the output of the $1^{st}$ round S-box for the CPA attack. The correct key byte is distinguished by the sharp spike in correlation ($\rho$) between the hypothetical HW leakage and the measured traces at the particular time instant where the target operation ($1^{st}$ round S-box) occurs for that key byte. 

In this work, we focus on one particular key byte ($13^{th}$ byte,  as it required the minimum number of traces to break for the unprotected implementation) to demonstrate the resiliency of our proposed countermeasure. 

\subsection{Motivation}
Although power analysis attacks have been known for more than two decades, the threat of power SCA is increasing with the growth of miniaturized and resource-constrained IoT devices. These small devices consist of low-power 8/16-bit microcontrollers which have high signal-to-noise ratio (SNR) making them more vulnerable to SCA attacks compared to the 64-bit processors (more `noise'). {\blue In case of 64 bit processor, algorithmic noise is more. For side channel analysis, attack model are built byte-wise or nibble-wise. Hence it has more chance to correlate to bytes being attacked for 8-bit/16 it architectures. On the other hand, parallel operations are done in 64-bit.} Hence, development of a low-overhead countermeasure is extremely critical to protect these embedded devices against power SCA attacks. 

In addition to low-overhead requirements, a countermeasure can be easily incorporated into a product if it is generic and synthesizable. Generic countermeasures are preferred from an industry standpoint as it helps to maintain legacy of the existing crypto algorithms and can be used as a wrapper without any modification to the crypto core. Synthesizable countermeasures helps in scalability across different technology nodes and does not require manual efforts, aiding to non-recurring engineering costs. These are important factors for an industry to adopt a particular countermeasure, and all of these have motivated the design of the proposed technique.

This article demonstrates a time-varying transfer function based low-overhead physical countermeasure utilizing switched capacitors to reduce the information content of the leakage from the crypto engine. Using time-varying transfer functions (TVTF) by efficient randomization of physical resources in the form of switched capacitors, the traces are significantly obfuscated, without any performance degradation This is a low-overhead circuit-level generic countermeasure and can be extended to any other crypto algorithms. Moreover, the circuit is entirely digital and can be synthesized (barring the capacitors), aiding technology scaling.

\subsection{Contribution}
The specific contributions of this work are:

\begin{itemize}
    \item This paper proposes physical time-varying transfer functions (TVTF) to obfuscate the leakage due to the crypto operations in the power traces. TVTF is achieved by efficient randomization of distributed switching capacitors.
    \item With the proposed TVTF, we mathematically and experimentally demonstrate the effect of multiple capacitors charging from supply/driving the AES at a given phase, revealing that randomly choosing a single capacitor each for charging/driving AES is the best choice to achieve the maximum power SCA protection.
    \item System-level simulation results in TSMC 65nm CMOS technology shows that the power SCA immunity is enhanced by $>4000\times$ compared to the unprotected implementation with only 20\%, 25\% area and power overheads respectively, and without any performance degradation. Moreover, the proposed countermeasure is generic and digital-friendly allowing scalability across different technologies.
    {\blue \item The paper proposes a solution and represents a mathematical model for it. All the components are assumed to be ideal to validate the immunity provided by our countermeasure with respect to the unprotected implementation. Practically, system noise would exist which will make the SNR of the power signatures even lower enhancing the MTD.} 
\end{itemize}

\subsection{Paper Organization}
The remainder of the paper is organized as follows. Section II discusses existing state-of-the-art in detail, along with the analysis of previously proposed switched capacitor based countermeasures. In section III, theoretical background and analysis of the proposed TVTF countermeasure is presented. Section IV discusses more experiments and shows a mathematical formulation to evaluate the efficacy of the proposed TVTF based multi-phase switched capacitor technique. Next, section V presents the implementation results, followed by {\blue section VI which analyses the tuning knobs for MTD improvement. Finally, section VII concludes the paper.}

\section{Background and Related Works}
\subsection{State-of-the-art Power SCA Countermeasures}
\begin{figure*}[!t]
\centering
\includegraphics[width=2\columnwidth]{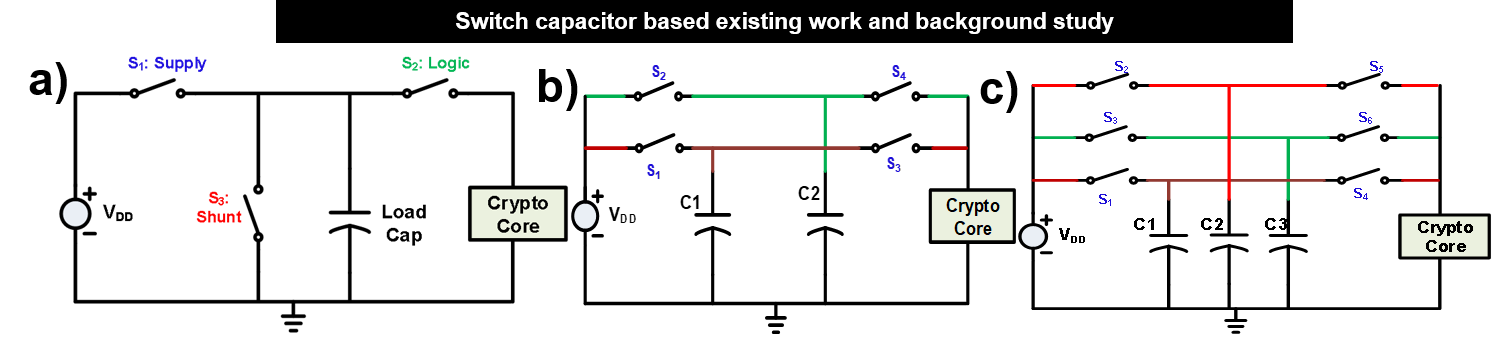}
\caption{\label{fig:carlos_archi} a) Switched Capacitor Current Equalizer Countermeasure proposed in \cite{tokunaga_secure_2009}, \cite{tokunaga_securing_2010}. It has 3 phases of operations: 1. charging load capacitor 2. charging crypto core from load capacitor 3. Reset the load capacitor voltage value to a predefined voltage value.  b) 2 phase switch capacitor without current equalizer solution. Switching activity explained in Table \ref{tab:archi_1_table}. Similar solution is proposed in \cite{Shamir2000ProtectingSC}. c) Multiple capacitor based circuit. Switching is mentioned at Table \ref{tab:archi_2_table}.}
\label{sw_cap_carlos}
\end{figure*}

The state-of-the-art hardware countermeasure for power SCA resistance can be broadly classified into three categories - logical, architectural and physical. The first category of logical countermeasures focus on designing SCA resistant logic styles to equalize the power in each cycle of the clock. This includes dual-rail precharge (DRP) logic style \cite{danger_overview_2009} and logic-level hiding like the sense amplifier based logic (SABL) \cite{tiri_dynamic_2002}, both of which require custom library cell design, and also incurs a large area overhead. Other logic-level hiding techniques like the wave dynamic differential logic (WDDL) \cite{hwang_aes-based_2006}, \cite{tiri_logic_2004}, and bridge boost logic (BBL) \cite{lu_1.32ghz_2015} also fall under this category, however, they are based on single rail standard cell libraries. WDDL was the first power SCA resistant circuit validated in silicon with a MTD of ~21K, incurring a $3\times$ area, $4\times$ power, overheads as well as $4\times$ performance degradation. Logic level masking at the gate level include masked dual-rail pre-charge logic \cite{popp_masked_2005}, \cite{popp_evaluation_2007}, which can be built using the standard library cells, however, it suffers from high area and power overheads.

The second category is architectural countermeasures, which can utilizes time or amplitude distortions to hide the leakage. Random insertion of dummy operations, shuffling of operations, clock randomization, random order execution fall in this category of architecture-level hiding. These shuffling techniques involving randomizing the order of instructions are limited by the number of instructions that can be shuffled depending on each algorithm, and does not provide high level of protection \cite{veyrat-charvillon_shuffling_2012}. Clock randomization based countermeasures including dynamic voltage and frequency scaling (DVFS) has been shown to be defeated by observing the clock edges at the supply \cite{baddam_evaluation_2007}. Masking schemes in the architecture level include boolean masking, masking multipliers and random pre-charge. All these countermeasures are typically algorithm and architecture specific and hence is not generic as it requires modifications in the algorithm itself.

The final category of power SCA protection is the physical countermeasures. The most well-known scheme in this category is noise injection. However, noise insertion suffers from extremely high area and power overheads \cite{guneysu_generic_2011}, \cite{das_asni:_2018}. Other techniques in this category are based on supply isolation. Low-dropout (LDO) regulators have been shown to provide power SCA resilience \cite{singh_25.3_2019}. However, it has also been shown that an ideal series  LDO implementation is inherently insecure \cite{das_asni:_2018}, \cite{das_high_2017}. Buck-converter based integrated voltage regulators (IVRs) suffer from area overhead due to embedded passives \cite{kar_8.1_2017}. {\blue An on chip signal suppression based countermeasure has been proposed in \cite{Ratanpal2004AnOS}.} Recently, Das et al. \cite{das_asni:_2018}, \cite{das_high_2017} proposed signature attenuation to enhance the minimum traces to disclosure (MTD) significantly. Although {\green signature suppression} is an efficient SNR reduction technique, {\green utilizes mixed-signal circuit} (high output impedance current source biased in saturation), which are not {\green easily} scalable across different technology nodes. {\green Hence, there is a strong need of synthesized physical generic low-overhead solution.}

Another physical-level countermeasure proposed by Tokunaga et al. utilizes a switched capacitor technique to isolate the AES engine from the power supply \cite{tokunaga_secure_2009}, \cite{tokunaga_securing_2010}. This is a novel circuit-level technique as it improves the MTD significantly ($>2500\times$) {\green but suffers from performance degradation}. Let us look into the operation of this circuit in the following sub-section. {\blue Further improvement has been discussed in \cite{moradi2016moments} to reduce crosstalk further which is claimed to be more immune to SCA attacks. }

\subsection{Switched Capacitor Current Equalizer Countermeasure}

The idea behind the switched capacitor current equalizer is to isolate the AES engine from the supply using a charging capacitor \cite{tokunaga_secure_2009}. This countermeasure, as shown in  Fig. ~\ref{sw_cap_carlos}a. has three phases of operation. In the first phase, switch $S_1$ is closed and the load capacitor gets charged. In the second phase of operation, switch $S_2$ is closed and the capacitor is connected to the crypto core, with complete isolation with the supply. In the third phase of this circuit (switch $S_3$ closed), the load capacitor is discharged (reset) to a fixed voltage so that the residual charge is not passed to the supply in the next phase (first phase: $S_1$ closed). 

To accommodate these three phases of operation, three identical switched capacitor modules are utilized providing uninterrupted AES operations. While this countermeasure provides high protection guarantees, it suffers from an $2\times$ throughput degradation. Iso-performance would require using high values of capacitances, leading to $>2\times$ area penalty. 

It needs to be noted that the third phase (reset) of operation is extremely important so that the discharged capacitor (connected to AES in second phase) is not directly connected to the power supply. 

{\green \textbf{Challenges of Reset Phase: }} The reset phase of the countermeasure involves a bias voltage which renders it non-synthesizable and would not scale across different technologies. {Also, every time resetting switch capacitors to a predefined voltage value increases voltage swing across the capacitor, which increments power overhead.} Hence, let us now try to leverage the switched capacitor based technique without the reset phase.

\subsection{Evaluation of Switched Capacitor Protection without Reset}


\begin{center}
\begin{table}[!t]
\caption{Switching Pattern for the 2-phase Switched capacitor without reset}
\centering
\begin{tabular}{ | c | c | c |}
\hline
Time instance & {Connected to $V_{DD}$}& Connected to AES\\\hline
 $t_{2n}$ & C1 & C2  \\ \hline
 $t_{2n+1}$ &C2 & C1 \\  \hline
\end{tabular}
\label{tab:archi_1_table}
\end{table}
\end{center}
\begin{center}
\begin{table}[!t]
\caption{Switching Pattern for the 3-phase Switched capacitor}
\centering
\begin{tabular}{ | c | p{0.8cm} | c | c |}
\hline
Time instance & \multicolumn{2}{|c|}{Connected to $V_{DD}$}& Connected to AES\\\hline
 $t_{6n}$ & \centering{C1} & C2  & C3 \\ \hline
 $t_{6n+1}$ & & & C1 \\  \hline
 $t_{6n+2}$ & \centering{C1} & C3 & C2 \\\hline
 $t_{6n+3}$ & & & C3 \\\hline
 $t_{6n+4}$ & \centering{C2} & C3 & C1 \\\hline
 $t_{6n+5}$ & & & C2 \\\hline
\end{tabular}
\label{tab:archi_2_table}
\end{table}
\end{center}
\begin{figure*}
\centering
\includegraphics[width=2\columnwidth]{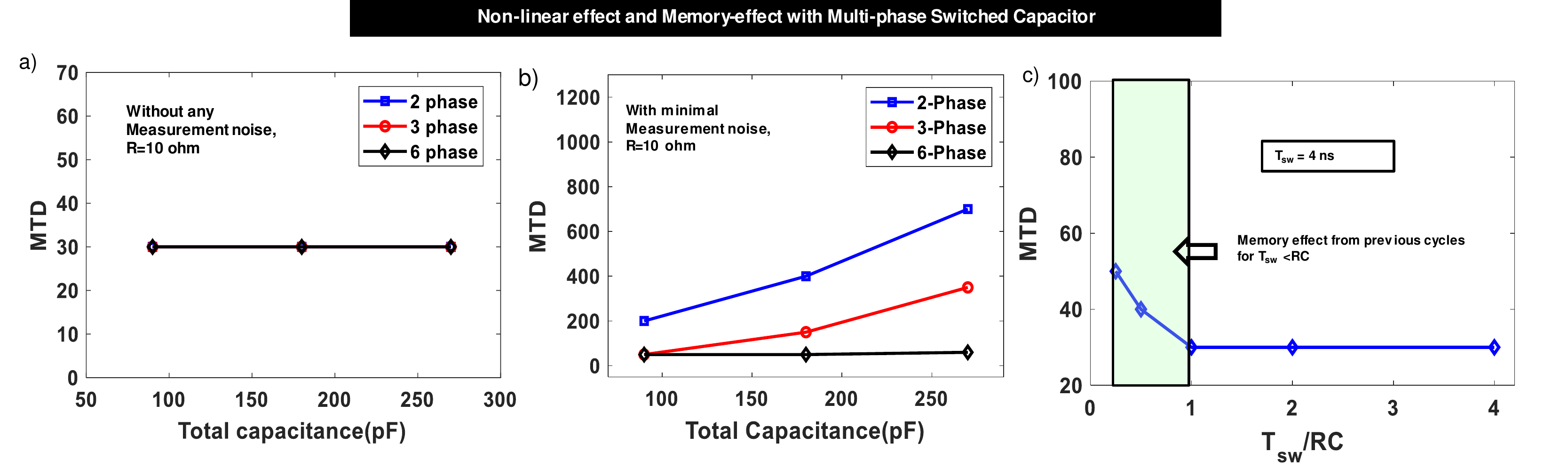}
\caption{\label{fig:nl_addition_wo_noise_w_noise} (a, b): Effect of residue voltage addition for multi-phase capacitor with and without measurement noise. Measurement noise is added to emulate the original noise present in the captured traces, to see the effect of signature attenuation due to the capacitor. It should be noted that this work focuses on physical time-domain shuffling and to analyze the efficacy of the proposed physical TVTF countermeasure, no additional measurement noise is added in the rest of the paper, unless mentioned otherwise. (c): Memory effect (shown for baseline 2-phase circuit) helps in cases of very high time constant (RC) or at higher switching frequencies. At our chosen RC (in the flat region, $C = 200pF$, $R=10\Omega$), MTD is not increased. Also in the zone where memory effect is useful, increasing switching frequency will increase power overheads, while high R, C choices will increase the area overhead.}
\end{figure*}
To make the switched capacitor current equalizer circuit synthesizable, we need to get rid of the third phase of operation, where the capacitors are getting reset to a fixed bias (analog) voltage. The modified circuit shown in Fig. \ref{sw_cap_carlos}b. consists of two phases with two load capacitors. In the first phase (t0), the capacitor (C1) is connected to the AES core, while the other capacitor (C2) is connected to the supply for charging. In the alternate phase (t1), C2 drives the crypto engine while C1 is charged from the supply.
The residual voltage on a capacitor after it has been been connected to the AES is given by,
\begin{equation}
    \label{eqn1}
    V_{res} = V_{DD} - \frac{1}{C} \int_{t_n}^{t_n+T} i_{AES} dt
\end{equation}
where $i_{AES}$, $T$ and $C$ are the AES current, switching period and capacitance of each unit capacitor respectively. Hence, the supply current as a function of time is given as,

\begin{equation}
    \label{eqn2}
    i_{sup}(t) = \frac{V_{DD}-V_{res}}{R} e^{-\frac{t}{RC}}
\end{equation}
where $R$ is the ON resistance of the switch. From Eqn. \ref{eqn2}, it is clear that the entire residue (integrated voltage) gets connected to the supply thereby leaking through the power supply. {\blue Similar approach has been taken in \cite{Shamir2000ProtectingSC}. In this work, capacitors have been included in packaging instead of IC, which makes it vulnerable to invasive attack. And, this isolation just changes the traces in a deterministic manner which means in case of CPA, correlation point will change though it will still correlate. Attenuation due to capacitors will slightly increase MTD.  }

It is interesting to note that we observe very small improvement in MTD ($<10\times$) with the 2-phase switched capacitor without reset compared to an unprotected implementation (initial MTD $\sim30$), since this circuit does not achieve any supply isolation. 

Next, we study the effect of multiple phases of the switched capacitors without reset and examine if addition of multiple phases which causes non-linear transformation to power trace has any significant role in providing SCA protection. 

\subsection{Multi-phase Switched Capacitor Implementation}


Here, we explore the effects of non-linearity (NL) and memory by charging multiple capacitors together in a phase, while another capacitor drives the AES engine in that phase. Fig. \ref{sw_cap_carlos}c. shows a three-phase switched capacitor circuit without any reset phase. Note that we refer to this circuit as three-phase because of the three capacitors (N = 3) which connect to the AES one at a time. Table \ref{tab:archi_2_table} shows the switching activity for the three capacitors. This strategy can be extended to larger number of distributed switching capacitors.

\subsubsection{Effect of Non-Linearity due to Multi-Cap Charging}

{\green Integration is a non-linear operation. Using capacitor integrates current trace over a specified time to introduce non-linearity. On the other hand, if those capacitors can be onnected to AES and VDD in different time, it can create time variance too.} Fig. \ref{fig:nl_addition_wo_noise_w_noise}(a) shows that the effect of \textcolor{black}{just introducing non-linearity} does not create any time-variance to enhance the MTD. Note that in Fig. \ref{fig:nl_addition_wo_noise_w_noise}(a), the effect of signature attenuation due to the load capacitor is not present, which is a simulation (modeling) artifact. As the AES power traces are collected from a real device and fed to the circuit simulator (Cadence Virtuoso) as a current piece-wise linear file (ipwlf), any amplitude reduction of the signature would also reduce the measurement noise in that signal, keeping SNR constant. This is an artifact for which previous works on signature attenuation \cite{das_asni:_2018} have also considered small noise injection to emulate the measurement noise initially present in the power traces. However Fig. \ref{fig:nl_addition_wo_noise_w_noise}(a) drives \textcolor{black} {the fact that increasing the number of phases does not have any impact in the time-domain obfuscation of power trace}. 

\begin{figure*}
\centering
\includegraphics[width=2\columnwidth]{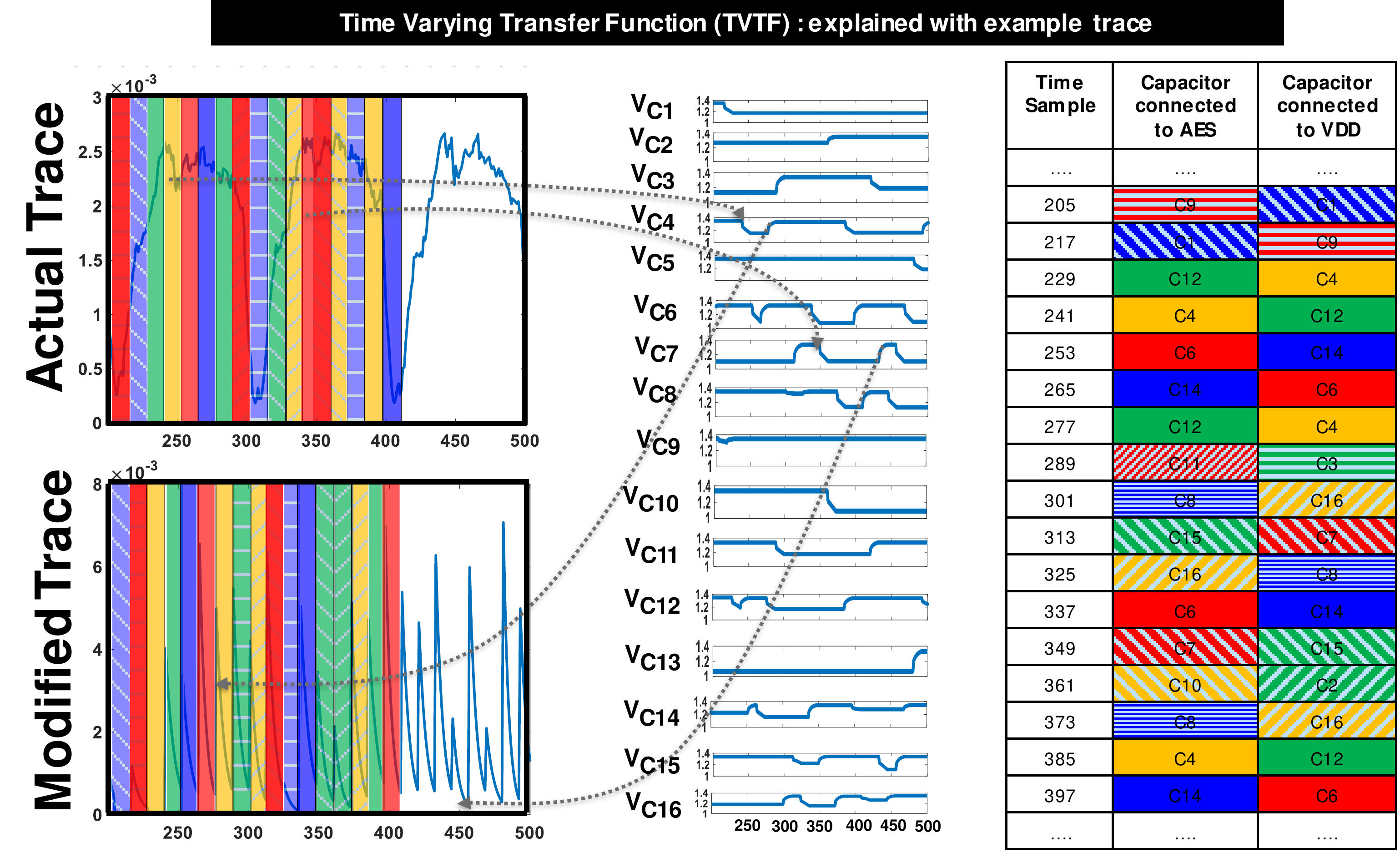}
\caption{This figure shows an example of TVTF-based randomization within a single cycle. AES trace is integrated and shuffled utilizing Algorithm I providing significant obfuscation in the modified power traces. It should be noted that this figure only shows obfuscation within one cycle, but in general, the obfuscation is not limited to a particular cycle and the information content of one cycle may be spread to few cycles later depending on the randomization algorithm, which chooses when a particular capacitor is charged from the supply.}
\label{fig:distortion}
\end{figure*}

Hence, any influence on the MTD is solely due to the signature attenuation as the voltage fluctuation across the AES gets suppressed by the load capacitor and part of it gets reflected through the ON switch during the charging phase. To observe this effect of signature attenuation, we inject a small amount of noise calculated from the initial SNR (20dB) of the captured traces. With the peak AES current of $3mA$, SNR of 20dB implies that the measurement noise present in the signal is $~0.3mA$. Emulating this measurment noise, we observe the effect of signature attenuation with the increase in total capacitance as shown in Fig. \ref{fig:nl_addition_wo_noise_w_noise}(b). Now, 2-phase (2 unit capacitors) switched capacitor implementation shows higher MTD than the 3-6-phase implementations as the unit capacitance becomes higher (total capacitance is constant for iso-area overhead). 

\subsubsection{Effect of Memory on MTD}
Next, we analyze the effect of memory of the distributed switched capacitor architecture on the MTD. After a capacitor has been connected to the AES engine, it has been discharged upto a certain voltage. \textcolor{black}{Now if we do not allow} it to charge back completely, i.e., if the switching period is much lower than the RC time constant ($T_{sw} < RC$, R being the ON resistance of the switch and C is the unit capacitance of the 2-phase switched capacitor), then the effect of previous samples can be spread across multiple next cycles, leading to power trace distortion. However, as seen from Fig. \ref{fig:nl_addition_wo_noise_w_noise}c, this obfuscation due to the memory effect is rather small and only increases the MTD slightly. Also, to leverage this small benefit would mean that the switching frequency ($f_{sw}$) is increased leading to a trade-off with the power overhead. {\green Another way to satisfy the condition is to increase either capacitor size or decreasing device size. (hence increasing impedance of the switches.) But, increasing capacitance highers area overhead. Decreasing device size beyond a point (length of the device according to different technology) is impossible and MTD does not increase much with respect to resistance of switches operating in linear region as shown in Fig. \ref{fig:nl_addition_wo_noise_w_noise}c.}
\textit{Hence, partial charging is not an efficient technique to enhance MTD as information is still being leaked despite some distortion in power trace.}

From these observations, we can conclude that multi-phase switched capacitor does not produce significant distortion of the power traces and can be broken easily. Definitely increasing the capacitance increases the signature suppression enhancing MTD at the cost of very high area overheads. Hence, our goal is to achieve high power SCA protection with low capacitances (low area overhead) and utilize physical time-based obfuscation techniques. 
It should be noted that for the rest of this work, we do not consider the effect of measurement noise, unless mentioned otherwise, as we focus on the evaluating the efficacy of physical time-domain {\green obfuscation}, rather than the effect of signature attenuation. 

This work utilizes the multi-phase distributed switched capacitor technique with physical time-domain pseudo-random obfuscation of the traces and demonstrate high SCA immunity with low area and power overheads and without any performance degradation.

\section{Multi-Phase Switched Capacitor with Physical Time-Varying Transfer Function}

In the previous sections, it has been shown that multi-phase capacitors in itself does not provide sufficient immunity to protect against power SCA attacks. Without the reset phase, the residual voltage of the capacitors leaks to the power supply and breaks within a small MTD. Now, if we can somehow randomize these multi-phase switching capacitors driving the crypto module such that they connect to the power supply at different points in time, the information content can be reduced significantly as the encryption traces become obfuscated across different points in time.

{\blue Fig. \ref{fig:distortion} shows an pictorial representation of the concept. Note that power traces will be available to the attacker in randomly obfuscated manner. } We implement a pseudo-random algorithm to determine the capacitor that is being charged at a time and also the one which drives the AES. This allows physical shuffling of the distributed load capacitors and obfuscates the traces across different time samples. {\green It is important to note that this is different from algorithm shiflling which has been introduced in multiple literature as it is done in VDD level with switch capacitor. Hence it is generic and easily applicable to any other crypto-algorithm.} Algorithm \ref{my_second_algo} presents physical TVTF technique for the randomized shuffling of the capacitors, by choosing only 1 capacitor (out of $n$ total capacitors - $n-phase$ switched capacitor implementation) to drive the AES and another to be charged from the supply at a particular time. Each clock cycle is divided into $n$ different phases and two capacitors are chosen by the algorithm, one for charging and the other drives the AES engine. 

\LinesNumbered
\begin{algorithm}[!b]
\label{my_second_algo}
\caption{Obfuscation Algorithm for TVTF}
  Take n number of capacitor.\\\
  Precharge the capacitors.\\\
Divide it in 2 different arrays . \\\ 
\While{Encryption is not done}{
Pick randomly 1 from 'to\_be\_charged' array and connect it to VDD.​\\\
Pick randomly 1 cap. From 'to\_supply\_AES' array and connect to AES.​\\\
After dt time put back those 2 capacitors in alternative arrays.\
}
\end{algorithm}

\begin{figure*}[!t]
\centering
\includegraphics[width=2\columnwidth]{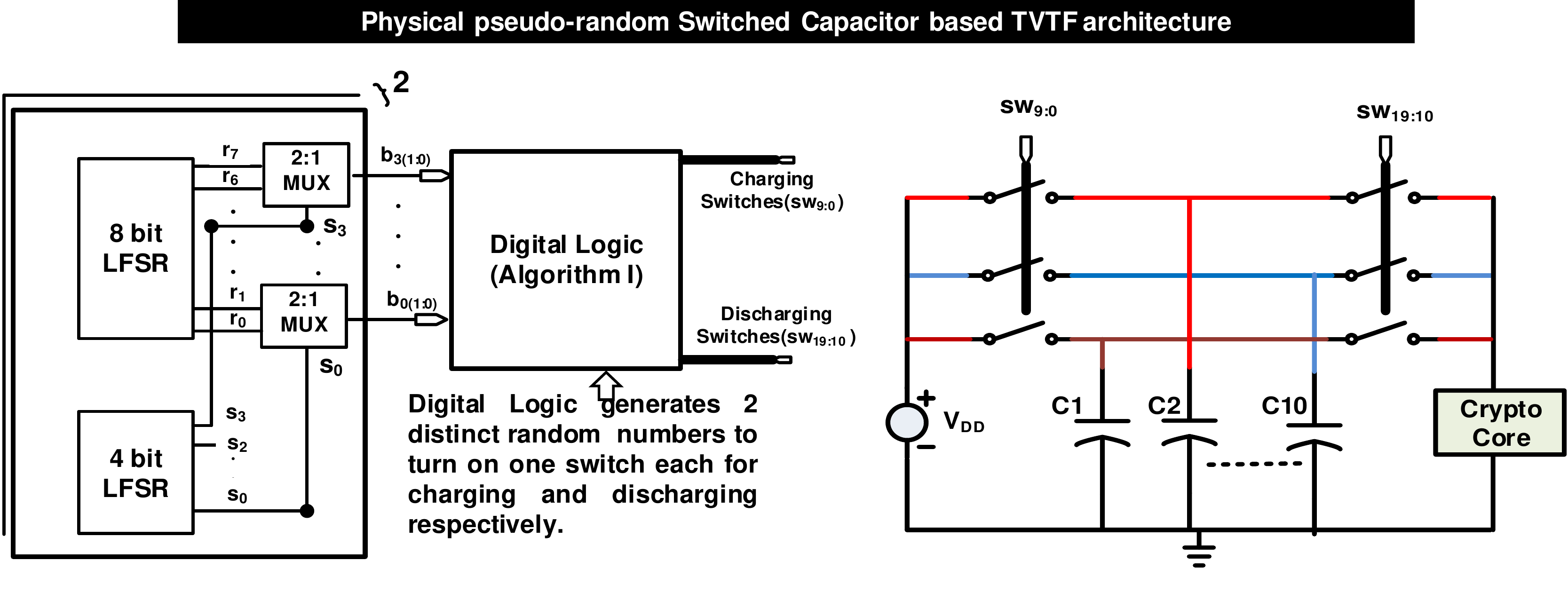}
\caption{Final Architecture of physical TVTF based multi-phase switched capacitor.}
\label{fig:archi_3} 
\end{figure*}

The algorithm is synthesized using hardware description language (HDL) by incorporating linear feedback shift registers (LFSR) as shown in Fig. \ref{fig:archi_3}. For the 10-phase TVTF switched capacitor implementation, we utilize 2-level stochastic LFSRs to obtain a high periodicity of $2^{8} - 1 = 255$. A 4-bit LFSR is used to stochastically sub-sample the 8-bit LFSR to produce a 4-bit output. {\green Similarly, we can increase the width of 1st LFSR and by appropriately sub-sampling it using proper multiplexer, we can get larger periodicity of output values. }

HDL version of Algorithm \ref{my_second_algo} takes two random number generated by both the LFSRs and ensures two numbers are different, so that a capacitor can not be connected to supply and AES at the same time. Hence, AES can never be connected to supply {\green directly}. This block also decodes the logic to turn on one switch for both the charging and discharging switches.  

This strategy of physical time-varying transfer function based shuffling obfuscates the signal drastically and reduces the information content as shown in Fig. \ref{fig:distortion}. If there are $n$ number of capacitors in the switched capacitor array, any random capacitor can be chosen in $n\choose1$ different possible ways. The probability of one particular capacitor to be charged at a time sample is $1 \over n-1$. \textbf{Hence, MTD will depend on the pattern of repetition of the shuffled capacitors and thus we employ 2-stage LFSRs (Fig. \ref{fig:archi_3}) to leverage high level of randomness.}

It should be noted that the power consumption of a LFSR is much lower compared to the AES itself. Hence, it will be very difficult for an attacker to retrieve the initial seed of the LFSR from the power traces. The seed can be programmed once by the manufacturer.{\blue After every operation done value of LFSR will be stored. This updated LFSR value will never be reset to initial seed, instead, updated value will act as seed for next operation. Thus, the chance of physical attack by collecting traces every time just after switching on the circuit and collecting the power traces.}

In this section, we have discussed the proposed TVTF approach by choosing one capacitor each ($n \choose 1$) for charging and discharging phase respectively out of the $n$-capacitor array. In the next section, we will evaluate the effect of choosing multiple capacitors ($m$) each for charging and discharging phases ($n \choose m$). {\blue Note that, the proposed architecture is fully synthesizable except the capacitors which have been used to store the charges for small amount of time and used as supply of crypto –engine (AES here). Control unit is all digital and through the power switch, it is connected to capacitors. Ports from control unit are available to make direct connection to the capacitors making the circuit digital technology scalable.}

\section{TVTF with Multi-Phase Switched Cap: Evaluation of $n \choose m$ Approach}

This section analyzes the effect of choosing multiple capacitors for charging and discharging at each phase of operation. As mentioned earlier, it should be noted that we divide the clock cycle into $n$ phases, where $n$ is the total number of capacitors in the switched capacitor array. 

We can choose $m$ capacitors out of the $n$ total capacitors in $n-m \choose m$ different ways. Now, we need to ensure that the charging of these $m$ capacitors through the supply does not overlap with the time points for another encryption. This combination can occur in $(n-m-m) \choose n$=$(n-2m) \choose m$ different ways. Even if one of the $m$ capacitors are being charged from the supply at the same time points, information gets leaked.  

Hence, the probability of information not leaking through the supply is given as, 

\begin{equation}
p_{not\_leak}\\={{n-2m \choose m} \over{n-m \choose m}} \\
\end{equation}

Hence, the probability of information leakage is given as ($P_{leak}$)

\begin{align}
\nonumber
    P_{leak}&= 1- {P_{not\_leak}}\nonumber
    \\&= 1- {{n-2m \choose m} \over{n-m \choose m}}\nonumber
    \\&={1-{{(n-2m)(n-2m-1)...(n-3m+1)} \over{(n-m)(n-m-1)...(n-2m+1)}}} 
\end{align}

Now ${{(n-2m)(n-2m-1)...(n-3m+1)} \over{(n-m)(n-m-1)...(n-2m+1)}} \leq {{n-2}\over{n-1}} \forall\ m $\\
${{n-2}\over{n-1}}$ is $P_{not\_leak}$ for m=1.
Equality condition exists for m=1. Hence for integer $m > 1$ probability of information leakage is more. 

Hence, with n-phase switched capacitor array, $n \choose 1$ ($m=1$) is the best possible TVTF strategy for physical time-domain obfuscation, as it reduces the information leakage by the maximum amount.

\section{Results : Design space exploration}

Power traces have been collected from an 8-bit Atmel microcontroller running AES-128 encryption, using the Chipwhisperer platform. The clock frequency of the software AES is 125MHz and has a peak current of $3mA$ (average current $\sim1mA$). A CPA attack performed on this unprotected AES-128 showed an MTD of $\sim20$ traces.

\subsection{Choice of Switch ON resistance \& Unit Capacitance}

\begin{figure}
\centering
\includegraphics[width=\columnwidth]{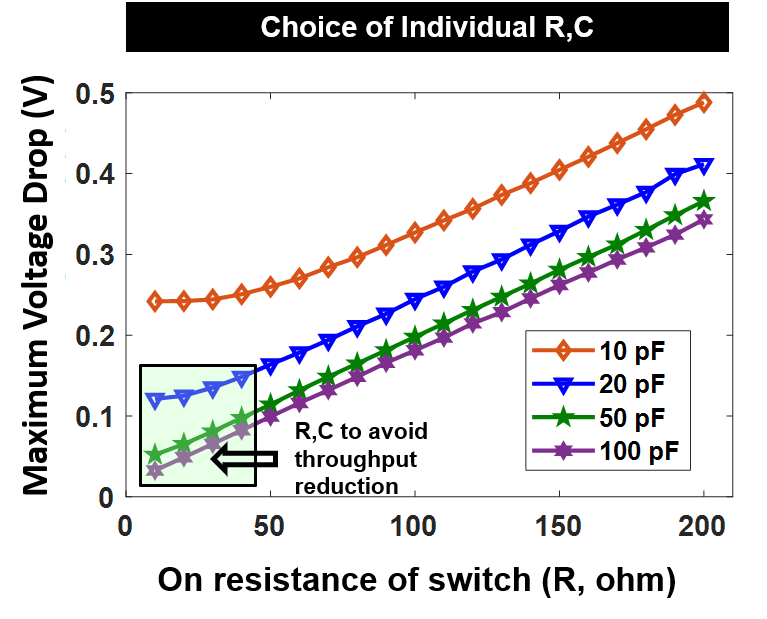}
\caption{Choice of the switch ON resistance (R) and the unit capacitance (C) is shown. For our TVTF-based switched capacitor circuit implementation, switch resistance (R) is chosen as $10\Omega$ and individual unit capacitors is chosen to be 20pF to avoid any performance degradation (voltage droop $< 100mV$).}
\label{fig:del_v_vs_rc} 
\end{figure}

As discussed earlier, to minimize the area overhead, a total capacitance of 200pF is chosen. Now we need to determine the optimal value of the unit capacitors such that there is no performance degradation of the crypto engine. Fig. \ref{fig:del_v_vs_rc} shows the effect of the switch ON resistance ($R$) and the choice of the unit capacitor ($C$) on the voltage droop across the AES core. Tolerating a maximum voltage drop of 0.1V, we see that with $R=10\Omega$, we can support a minimum unit capacitance of 20pF. This also implies that the maximum limit to our number of phases/capacitors ($n$) becomes 10. 
\subsection{Effect of increasing the number of Phases}
Figure \ref{fig:MTD_vs_phi} shows that increasing the number of unit capacitors, and hence the phases of operation in a clock cycle increase the MTD significantly. Also, Figure \ref{fig:MTD_vs_phi} shows that the effect of memory due to the previous time cycles ($T_{sw} < RC$) has negligible effect on MTD. We also see that 10 phases (or capacitors) gives a higher MTD, and hence we choose 10 as the number of optimal phases. It should be noted that increasing the number of phases has trade-off with the power consumption (due to higher switching frequencies) and performance.
\begin{figure}
\centering
\includegraphics[width=\columnwidth]{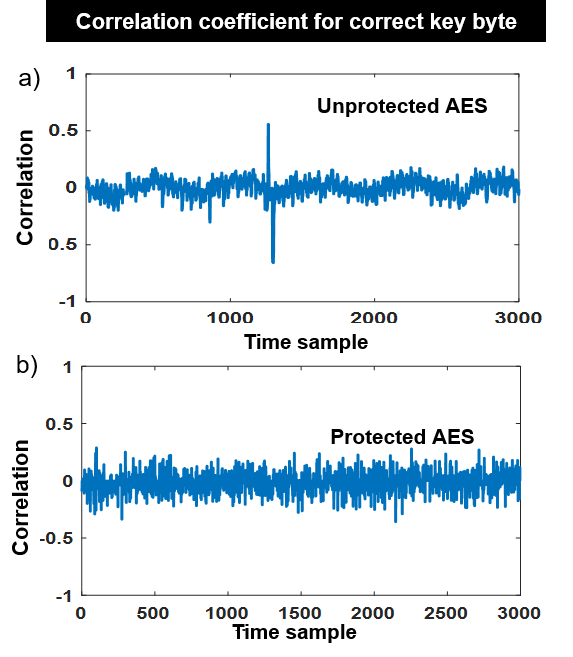}
\caption{a) Correlation between Hamming weight matrix and current traces after 200 traces for unprotected AES. b) Correlation between Hamming weight matrix and protected current traces for TVTF-AES.}
\label{fig:ro_vs_sample}
\end{figure}

As discussed in section III, next set of results are obtained by choosing multiple capacitor from an array of 10 capacitors. With increase of number of capacitors chosen at a time(m) reduces MTD. Even though we might expect a higher degree of randomization due to multiple capacitors getting charged/discharged, we see the opposite trend, which is consistent with our analysis in Section III. We see that for $n = 10, m = 1$ gives the maximum MTD, as seen from Figure \ref{fig:MTD_vs_10cm1}. This also reveals that TVTF allows for an area-efficient solution with very low values of unit capacitor. Although low values of capacitor has trade-offs with the performance degradation, as discussed earlier. 
\begin{figure}
\centering
\includegraphics[width=\columnwidth]{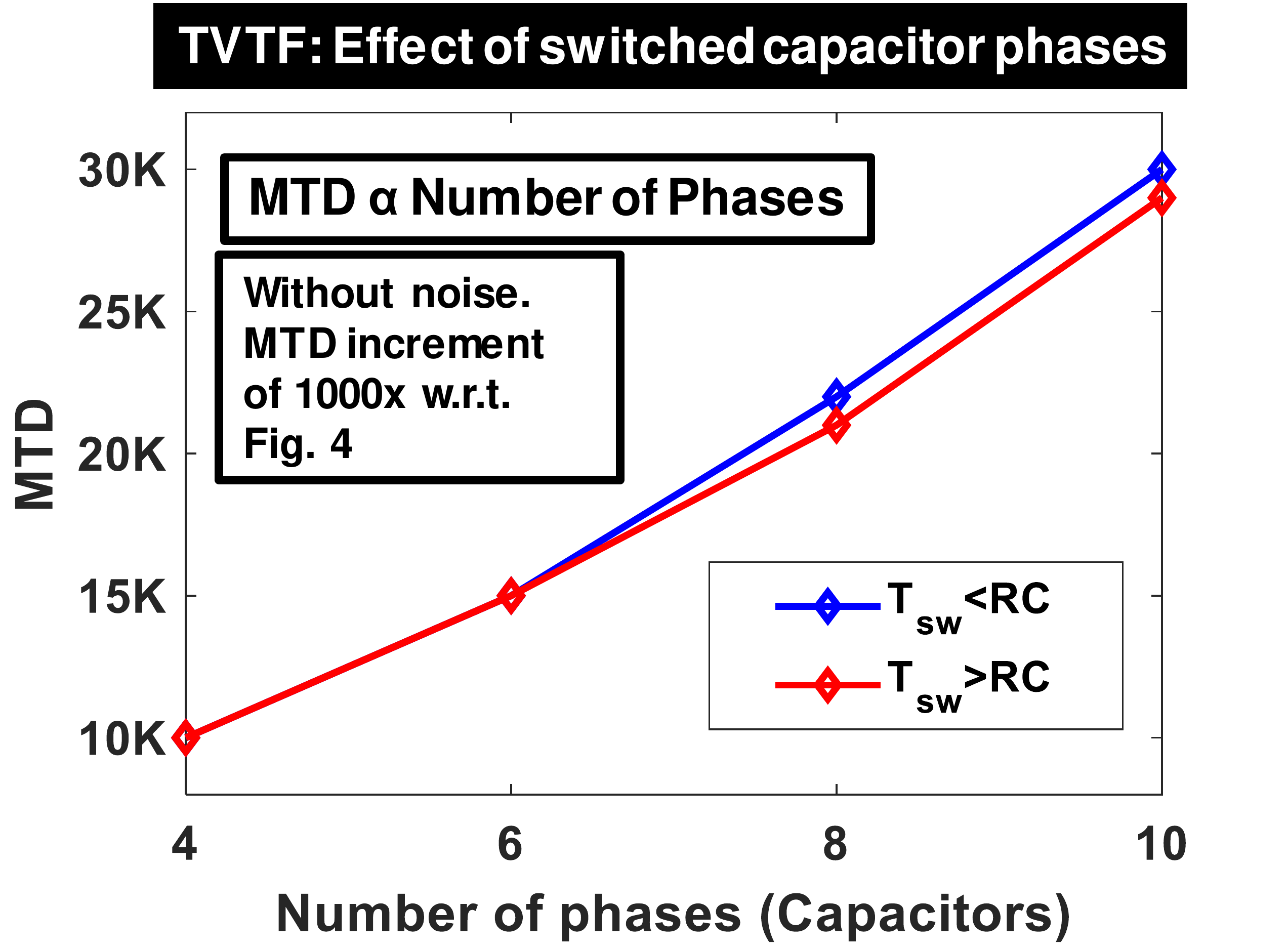}
\caption{MTD increases as the number of phases (unit capacitors) are increased. The memory effect for $T_{sw} < RC$ is quite small and does not justify the associated power overhead trade-off. Hence, the proposed TVTF circuit operates in the region of $T_{sw} > RC$.}
\label{fig:MTD_vs_phi}
\end{figure}
Figure \ref{fig:ro_vs_sample}(a) shows peak in correlation for the correct key with only 200 power traces, while the protected implementation does not show any correlation spikes (Figure \ref{fig:ro_vs_sample}(b)).
\subsection{Effect of Choosing multiple capacitors ($n \choose m$) with TVTF}
\begin{figure*}
\centering
\includegraphics[width=2\columnwidth]{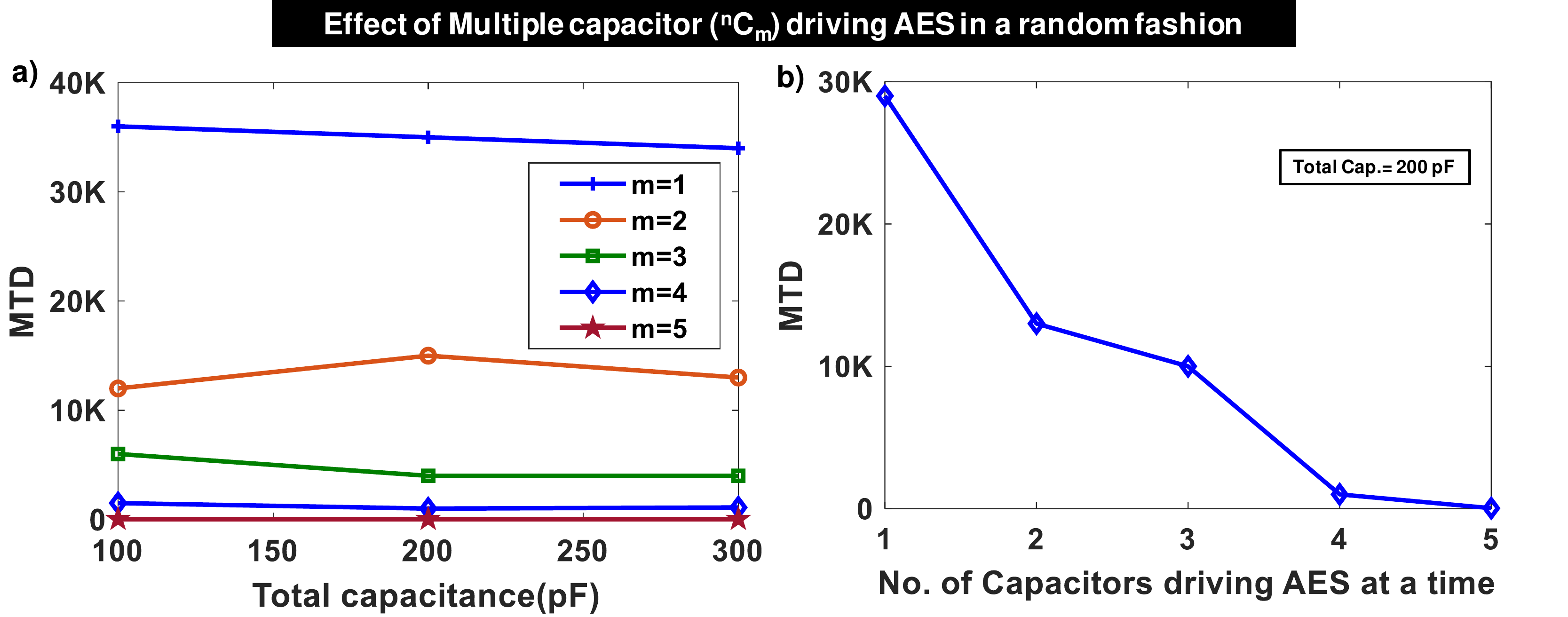}
\caption{a)Effect of the choice of number of capacitors (m) at a time from a pool of 10 capacitors (n = 10) on MTD. As discussed previously, increasing m does not enhance MTD as the probability of information leakage is increased. This also implies that a low value of unit capacitor is sufficient to leverage the benefits of TVTF-based time shuffling, but it has trade-offs with the throughput/performance. b)Effect of multiple capacitors for a fixed total cap (200pF) shows that the MTD is maximum when a single capacitor (m=1) is chosen by the TVTF algorithm, each for the charging and discharging. The decreasing trend with respect to $m$ supports the mathematical justification in Section IV that the $n \choose 1$ is a better randomization than $n \choose m$ for $m\ge2$.}
\label{fig:MTD_vs_10cm1}
\end{figure*}
\begin{figure*}
\centering
\includegraphics[width=2.1\columnwidth]{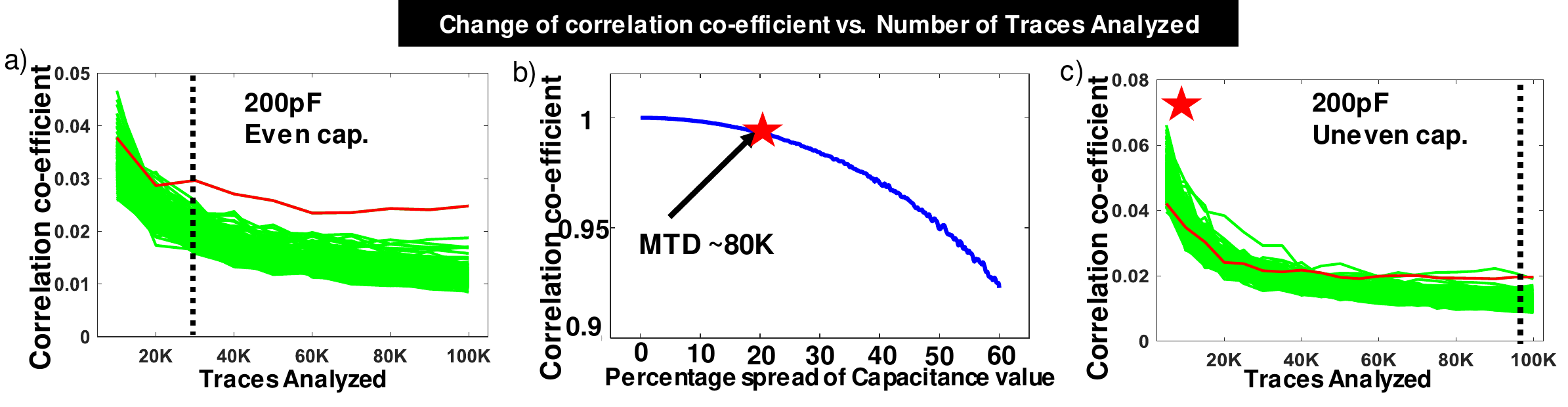}
\caption{a) MTD plot for the TVTF-based switched capacitor technique shows an MTD $\sim30K$ ($1500\times$ improvement) with equally distributed unit capacitors (20pF each) across 10 phases. b) With unequal capacitance ranging from 16pF-24pF with a total capacitance of 200pF produces protection up to MTD $>80K$ traces which is  $>4000\times$ compared to the unprotected AES implementation.}
\label{fig:MTD_vs_del_v}
\end{figure*}

Finally, Fig. \ref{fig:MTD_vs_del_v} shows the MTD plots with respect to number of traces analyzed for the proposed solution. For the TVTF based switched capacitor with 10 evenly distributed capacitors (n = 10, m = 1), we achieve an MTD of $\sim30K$ traces (Fig. \ref{fig:MTD_vs_del_v}(a)).

Next, we study the effects of uneven distribution of unit capacitors and also covering 2 clock cycles to achieve higher levels of randomization.
\subsection{Effect of Periodicity of PRNG}
Pseudo random number generator is the backbone of TVTF architecture. It has been observed with increase of periodicity, MTD has been increased significantly. This observation is tabulated in Table \ref{tab:period_table}. Note that, periodicity changes inversely effect the probability of getting same trace at same time point, which increases MTD.  
\begin{center}
\begin{table}[!t]
\caption{MTD improvement by tuning Periodicity of PRNG (Number of capacitors and total capacitance are fixed )}
\centering{
\begin{tabular}{ | c | c |}
\hline
{Periodicity} & {MTD} \\ \hline
{$2^{3} - 1$} & {700} \\ \hline
{$2^{16} - 1$} & {66000} \\  \hline
{$2^{32} - 1$} & {92000} \\  \hline
 \end{tabular}
 }
\label{tab:period_table}
\end{table}
\end{center}
\subsection{Effect of unequal capacitors}
Further randomization can be achieved with unequal capacitance values, while maintaining a fixed total capacitance of 200pF. With this, MTD can be further enhanced (Fig. \ref{fig:MTD_vs_del_v}(c). This is because according to equation \ref{eqn1} as the voltage residue value depends on individual capacitance. Hence presence of different capacitors further distorts the signal and increases protection by $2.5\times$ and increases the MTD to 80K traces ($4000\times$) with iso-area overhead.

{\blue

Voltage sample at $n^{th}$ time sample will be obfuscated according to proposed algorithm and will be available at different time sample for different cycle as shown in Fig. \ref{fig:uneven_cap_concept}. But we see from equation \ref{eqn1} capacitance value can further distort voltage trace as it is the co-efficient of current integration term. Hence introducing different capacitance value will lead to MTD increment. 
\begin{figure}
\includegraphics[width=\columnwidth]{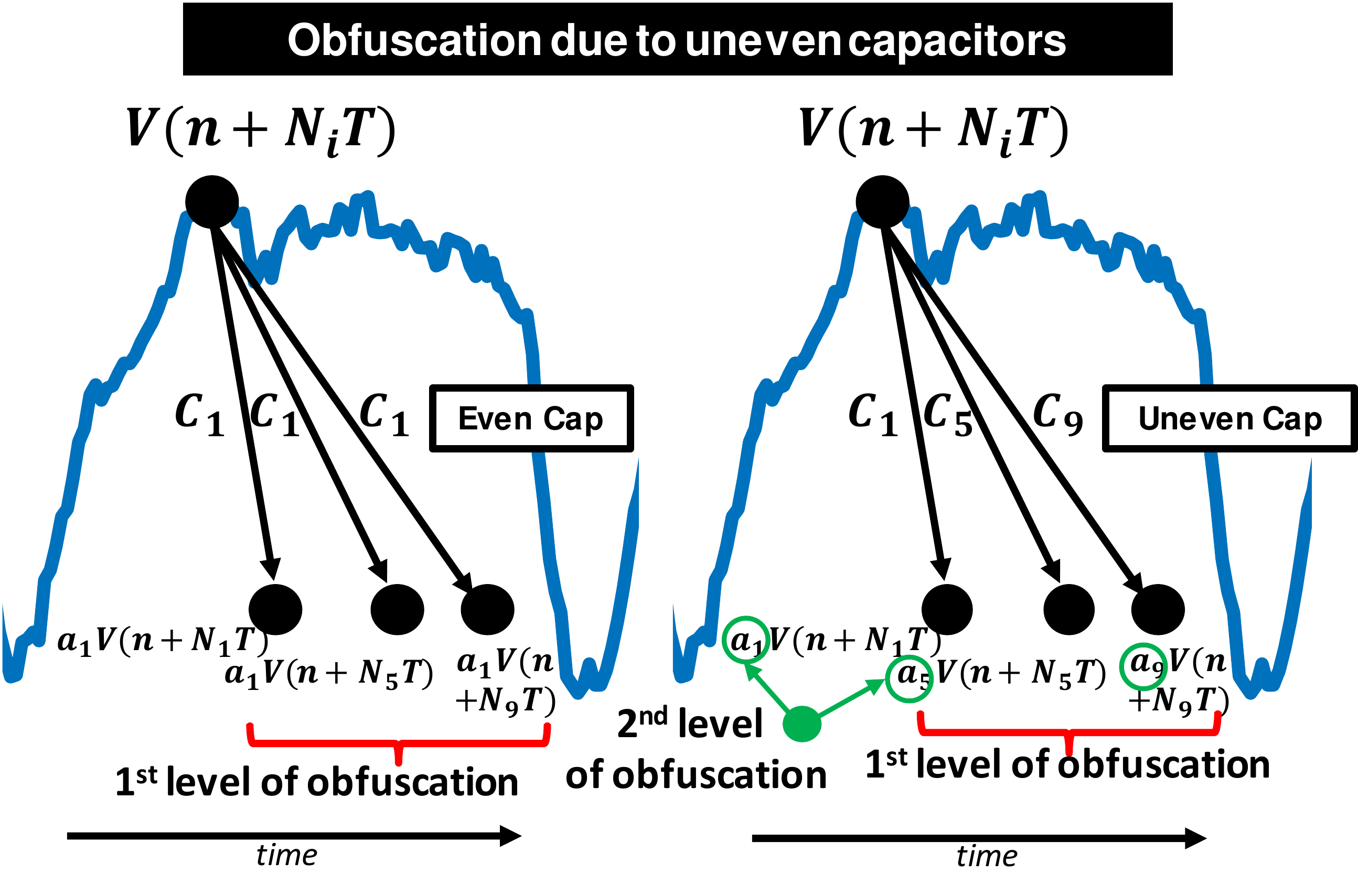}
\caption{It is shown in Equation \ref{eqn1} voltage residue depends on capacitance value of each capacitor. Introducing unevenness in capacitance adds an extra level of randomization.}
\label{fig:uneven_cap_concept}
\end{figure}

To check the sole effect of spread vector due to uneven capacitance value, we take a sample signal. We scale the signal according to effect of vector due to spread of capacitance value and correlate it with our initial signal. We observe change in correlation co-efficient as shown in Fig. \ref{fig:MTD_vs_del_v}b. Note that with 20\% in capacitance value spread gives a significant improvement in MTD (~4000x as shown in Fig. \ref{fig:MTD_vs_del_v})c. We can leverage this constraint to reduce correlation value even more which will increase MTD number significantly. Remember that, increasing spread of capacitance value implies reducing the minimum capacitance value which will increase maximum droop. It can adversely effect the efficiency of the circuit. To be on safer side, average capacitance has to be increased. Hence area overhead will be increased.   
}

\textbf{Effect of multi-cycle operation : }
We also analyze the effect of operating across multiple cycles with higher total capacitances and the same unit capacitance. Instead of covering 1 cycle with distributed capacitor, more cycles can be covered with increasing number of phases which will definitely increase order of randomization and increase MTD). This has some drawbacks too. Capacitances of distributed units can not be reduced significantly, as it increases the voltage drop the AES leading to performance degradation. Hence, in this case most obvious solution is to increase the value of capacitors, which again cause the area overhead.

\begin{figure}
\includegraphics[width=\columnwidth]{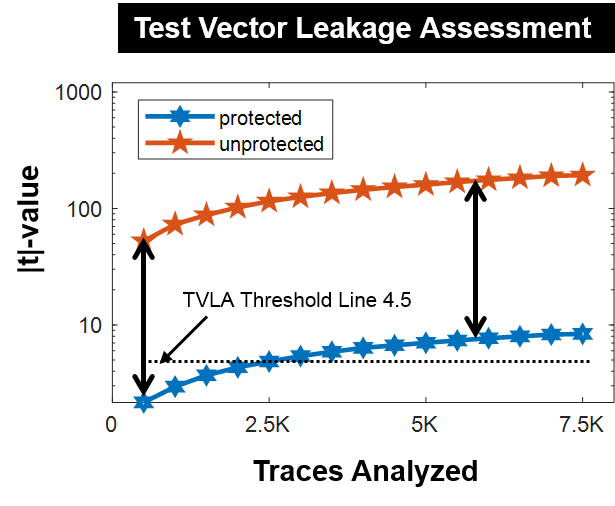}
\caption{Test vector leakage assessment results show significant amount of decrease in statistical t-value in protected version which implies encryption engine with proposed countermeasure is way less leaky with respect to straightforward implementation.}
\label{fig:tvla_after_7.5k}
\end{figure}

\begin{figure}
\centering
\includegraphics[width=\columnwidth]{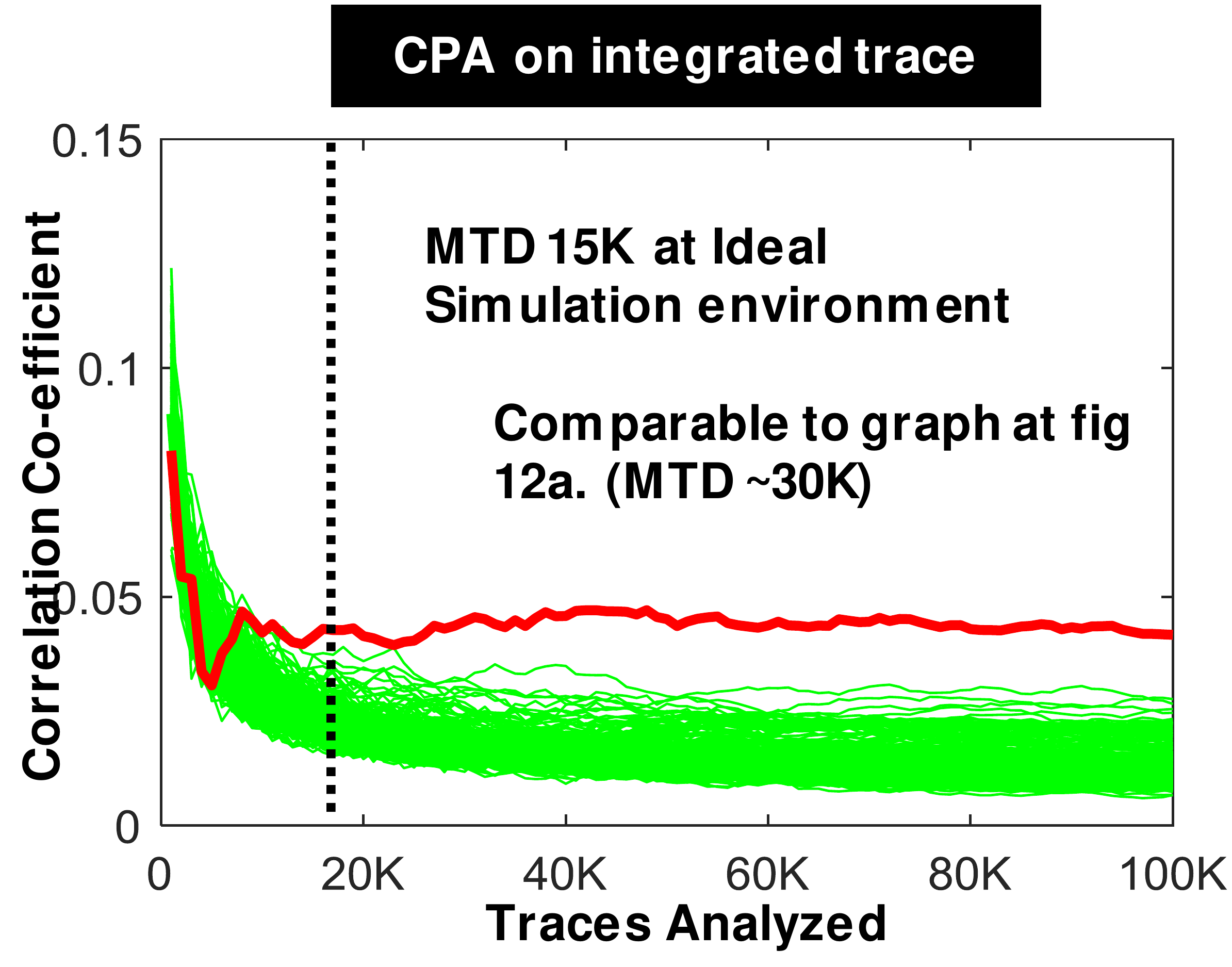}
\caption{Correlational power attack on integrated trace has shown an improvement over normal CPA. At ideal situation, MTD is 15000 in protected version. }
\label{fig:integration_attack_result}
\end{figure}

{\blue
\subsection{Test vector leakage assessment}
Surely, attacking using CPA is an indication but does not completely declare extra security. Different type of power analysis attacks have been introduced and researches are going on for further improvement on attack models and algorithms. Hence, it is important to calculate amount of meaningful leakage by an encryption engine. Test vector leakage assessment (TVLA) is one of the most trusted leakage assessment algorithm. $|t|-value $ of 4.5 does not have any data dependent leakage. AES with TVTF crosses the threshold of 4.5 after 2.5K traces while unprotected has a much higher value (11.5) even with only 5 traces. We observe maximum $|t|-value $ of 8.37 in the protected AES version against 190.1 of unprotected version after 7.5K traces. Trend has been shown in Fig. \ref{fig:tvla_after_7.5k}. 

\subsection{Immunity against correlational power analysis on sliding window based integrated trace}
Traditional CPA assumes a single leakage point and expects it to be correlated with attack model. Crypto algorithms (specially in software implementations) might have multiple leakage points which is usually located near each other \cite{Fledel2018SlidingWindowCA}. To test the immunity of our countermeasure, it has attacked using this algorithm. Proposed countermeasure has an MTD of 15000 at ideal simulation. MTD has been further increased to 97000 in presence of slight measurement noise as shown in Fig. \ref{fig:integration_attack_result}. It has been observed that software AES trace has a tendency to correlate at multiple points. These correlation points normally stay close to each other. Hence in integrated trace, this increases chance of getting correlated. This will not necessarily be better in other type of implementation where only one correlation point exists. 
}





\begin{table*}[!t]
\begin{center}
\caption{Overhead comparison of Time Varying Transfer Function (TVTF) with the existing state-of-the-art countermeasures }
\centering
\begin{tabular}{| p{0.20\linewidth} | c | c | c |c| c | c | c |}
\hline
Parameters & This Work & TCAS-I'18\cite{das_asni:_2018}  & JSSC '06\cite{hwang_aes-based_2006} & ISSCC '09\cite{tokunaga_secure_2009} & ISSCC '17\cite{kar_8.1_2017}  & ISSCC '19\cite{singh_25.3_2019}\\\hline\hline
 Technology & 65nm & 130nm  & 180nm & 130nm & 130nm & 130nm  \\\hline
 Technique used & TVTF & ASNI & WDDL & Sw. Capacitor & IVR & Digital LDO\\  \hline
 Power &1.24x  &1.68x & 4x & 2.66x & 2x &1.35x \\\hline
 Area & 1.2x &1.6x & 3x & 1.25x &2x &1.38x{\footnotemark}\\\hline
 Performance Degradation & 0 & 0 & 4x & 2x &0&1.1x\\\hline
 MTD & 4000x & 1000x & 30x & 2500x &100x&4210x\\\hline
 Comments & Digital-friendly & Mixed-signal & Mixed-signal& Mixed-signal&mixed-signal&Digital-friendly \\\hline
\end{tabular}
\end{center}
 \ \ \ \footnotesize{{\textsuperscript{1}Large metal-insulator-metal (MIM) load capacitor (1.9nF) not considered in the area overheads.}}
\label{tab:comparison}
\end{table*}
{\green Table IV compares this work with existing solutions. WDDL \cite{hwang_aes-based_2006} suffers from high overheads and performance degradation. Switch capacitor current equalizer circuit by Carlos et. al \cite{tokunaga_secure_2009} also suffers from performance degradation and is a mixed-signal circuit. IVR \cite{kar_8.1_2017} based countermeasure does not increase MTD to a large number. On a different note, Digital LDO based countermeasure \cite{singh_25.3_2019} has higher area overhead with capacitor. Proposed countermeasure has an area-overhead of ~4\% without the capacitors. }
\subsection{Area and Performance analysis}

Power overhead depends on 3 components. \\
\indent 1. Power lost while charging the capacitors and have been given by, \\$P_{l} = 0.5 \times f_{switching} \times unit\_capacitance \times (\delta V)^2 =0.125\ mW$,where\ $\delta V$ is the maximum voltage drop. \\
\indent 2. Another component is the switching power, $P_{switching}\\=f_{switching}\times C_{gate\_cap} \times V_{DD}^2\ =50\ uW,$ where\  $C_{gate\_cap}$\ is \ the\ gate capacitance \ of \ switch. \\
\indent 3. Pseudo-random number generator (PRNG) is another cause of power overhead. For a 10-bit LFSR, the power overhead is given by $P_{PRNG}=P_{LFSR}\times 2+P_{logic}=150\mu W$.
\\Final power overhead, \\
$P_{ov} = P_l+P_{PRNG}+P_{switching}=325\mu W$.\\
Hence, the power overhead can be given as,\\ ${{P_{AES}+P_{ov}}\over{P_{AES}}}={{0.325+1.32}\over{1.32}}=1.24 \times$.


Similarly Area overhead will have 3 components - area due to capacitors ($A_{cap}$), area of PRNG ($A_{PRNG}$) and area of the PMOS switch ($A_{sw}$).
Hence, area overhead is given as, $A_{ov}=A_{cap}+A_{PRNG}+A_{sw}=0.03\ mm^2$.\\

Area of AES in 65nm TSMC CMOS technology is $\approx 0.15$ mm\textsuperscript{2}.
Hence, the relative area overhead $= {{{A_{ov}+A_{AES}}\over A_{AES}}}\\={{0.15+0.03}\over{0.15}}=1.2\times$.\\
Table III summarises the MTD improvements and overhead comparison with respect to the existing state of the art countermeasures. 

\subsection{Remarks}
Power overhead linearly increases with higher switching frequency. At switching frequency of 1.25GHz, the power consumption becomes $325\mu W$.
Again from Fig. 10, it is evident that increasing number of phases per clock cycle of AES increases MTD, hence providing more immunity. Clearly, we can infer from these two trends that the number of phases can be used as a tuning knob to optimize between MTD and the power efficiency. 

Note that, in this work, we have chosen captured traces from a software AES running on an 8-bit Atmel microcontroller, and performed system-level simulations in Cadence Virtuoso. Although this a physical circuit-level countermeasure, it can be used as a wrapper both for hardware as well as software implementations of AES. The primary reason of choosing a software AES was to deal with low initial MTD values which helps in faster analysis of the proposed countermeasure.

{\blue Note that this paper mainly focuses on power side channel attack. Though EM side channel attack is also a threat to consider, it is beyond the scope of this work. But, this solution can be extended to EM Side channel attack too. Analysis has been shown in \cite{das2019stellar} that low level metal layers radiates very less amount of leakage from the IC. Hence capacitors shuffling logic as well as AES charging logic can be implemented using low level metal layer before it routes to highly radiating metal layers for charging of supply capacitors. EM probe will detect shuffled traces in this solution which is already immune to side channel attack.}

{\blue 
\section{Tuning knob to increase the resistance of proposed countermeasure}
This section summarizes the key factors that allows to increase immunity of proposed countermeasure even more at the cost of area or power overhead.
\begin{itemize}
    \item \textbf{Number of Phases:} Number of phases is one of the most important parameter to increase MTD. Number of phases implies increase in randomization within a given time window. Hence, probability of getting same traces at a particular point reduces further which helps to increase MTD. Fig. \ref{fig:MTD_vs_phi} shows the trend. Note that increasing number of phases requires higher switching frequency producing higher power overhead.
    \item \textbf{Effect of uneven capacitors: } Residue trace depends on the capacitance value of capacitors as discussed previously. Changing the capacitance of each capacitor and making it slightly uneven for each other further distorts the trace thus reducing the information.  
    \item \textbf{Periodicity of PRNG: } Periodicity of PRNG implies repetitive pattern in capacitor connection to VDD and AES. Hence increment in periodicity increases random shuffling and helps in MTD increment.
    
\end{itemize}

}

\section{Conclusions}
Power side-channel attack is a prominent attack on encryption ICs. This works proposes TVTF: a physical time-varying transfer function countermeasure to significantly obfuscate the power traces in the time-domain utilizing multi-phase switched capacitors. TVTF performs efficient randomization of the switched capacitors to obfuscate the traces. 

Previously, shuffling based architectural countermeasures have been proposed which randomize the order of instructions, but there are limited number of instructions which can be shuffled and are specific to a particular algorithm and architecture. DVFS-based countermeasures based on clock randomization were shown to be broken previously by observing the clock edges at the power supply, since it preserves the order of the instructions. 

The proposed TVTF-based switched capacitor countermeasure provides a generic, low-overhead ($1.2\times$ area, $1.25\times$ power overhead), and digital-friendly solution. The capacitors are not synthesizable, but the rest of the countermeasure is completely digital which makes it scalable across different technologies. Finally, it achieves a power SCA protection of $>4000\times$ compared to the unprotected implementation without any performance degradation. 

\bibliographystyle{unsrt}
{
    \bibliography{main.bib}
}

\end{document}